\documentclass[final]{cvpr}

\usepackage{times}
\usepackage{epsfig}
\usepackage{graphicx}
\usepackage{amsmath}
\usepackage{amssymb}
\usepackage{subfigure}
\usepackage{overpic}

\usepackage{bm}
\usepackage{algorithm}

\usepackage{algorithmicx}
\usepackage{algpseudocode}

\usepackage{enumitem}
\setenumerate[1]{itemsep=0pt,partopsep=0pt,parsep=\parskip,topsep=5pt}
\setitemize[1]{itemsep=0pt,partopsep=0pt,parsep=\parskip,topsep=5pt}
\setdescription{itemsep=0pt,partopsep=0pt,parsep=\parskip,topsep=5pt}

\usepackage[pagebackref=true,breaklinks=true,colorlinks,bookmarks=false]{hyperref}

\def\cvprPaperID{159} 
\def\confYear{CVPR 2020}

\graphicspath{{figures/}}

\begin{document}


\title{Morphological Anti-Aliasing Method for Boundary Slope Prediction}

\author{Yuchen Zhong ~~~~~ Yuchi Huo ~~~~ Rui Wang \\
}

\maketitle

\begin{abstract}
Image pixel aliasing caused by insufficient sampling is a long-standing problem in the field of computer graphics. It has always been the goal of researchers to seek anti-aliasing algorithms with high speed and good effect. Due to the deficiencies in local detection and reconstruction of sloping line boundaries, a morphological anti-aliasing method for boundary slope prediction is proposed. This method uses the information of the local line boundary slope to predict and test the end positions of the line boundary in the global scope, thereby reconstructing The boundary information more consistent with the actual boundary is obtained, and a more accurate linear boundary shape is obtained with only a small increase in the amount of calculation. Compared with the previous morphological anti-aliasing algorithm, the proposed method is based on the global morphological boundary. , can reconstruct the straight line boundary more accurately, and apply it to the anti-aliasing calculation, which can further improve the color transition of the straight line boundary, make the inclined straight line boundary have higher continuity, and obtain a better anti-aliasing effect.
\end{abstract}

\section{Introduction}

Image pixel aliasing caused by insufficient sampling is a long-standing problem in the field of computer graphics, which will cause still image distortion and dynamic image flicker. Directly increasing the number of samples can solve the aliasing problem, but the computational cost that increases linearly with the number of samples limits the The use of this method in real-time rendering. The main goal of current anti-aliasing technology research is to find an anti-aliasing algorithm with high speed and good effect.

Traditionally, supersampling anti-aliasing (SSAA) and multiple sampling anti-aliasing (MSAA) are the most widely used anti-aliasing algorithms in recent decades. However, simply increasing the number of samples is costly. Jorge et al. In the paper \cite{r1}, it is pointed out that the MSAA algorithm has a large memory consumption and GPU occupancy rate, and at the same time, multi-sampling is also difficult to combine with the popular deferred rendering (according to the test of Jorge et al., the GPU occupancy of the multi-sampling algorithm The rate can even be as high as $30\%$ in the framework of deferred rendering).

In recent years, Reshetov proposed a morphological anti-aliasing method \cite{r2}. Different from super-sampling anti-aliasing (SSAA) and multi-sampling anti-aliasing (MSAA), this method is an image post-processing anti-aliasing method. The boundary of the image in the frame buffer is detected, the original shape of the reconstructed boundary is inferred according to the different line types of the boundary, and the anti-aliasing calculation of the boundary pixels is realized by using the reconstructed boundary shape. , a good anti-aliasing effect can also be obtained by reconstructing the geometric boundary. However, the morphological anti-aliasing algorithm has its inherent defects: due to the lack of pixel sampling, the original geometric boundary information of the object has been lost on the image. The boundary information obtained by morphological detection is based on the local line type, and there is inevitably a certain deviation from the original boundary, which affects the quality of anti-aliasing. In order to further improve the effect of anti-aliasing, the morphological anti-aliasing algorithm of Reshetov \cite{r2} After the proposal, researchers have proposed a variety of post-processing anti-aliasing techniques based on image morphology detection, including fast approximate anti-aliasing (FXAA) \cite{r3}, distance-to-edge anti-aliasing (DEAA) \cite{r4}, directional The localized anti-aliasing (DLAA) \cite{r5}, sub-pixel reduction anti-aliasing (SRAA) \cite{r6}, sub-pixel modified morphological anti-aliasing (SMAA) \cite{r1} and so on. These algorithms are used in area calculation, sub-pixel feature The original morphological anti-aliasing algorithms have been improved in many aspects. However, these algorithms still have defects in solving the deviation of the morphologically reconstructed boundary from the original boundary. High-quality anti-aliasing of sloping straight-line boundaries is still an urgent problem to be studied. question.

\begin{figure*}
    \centering
    \includegraphics[width=\textwidth]{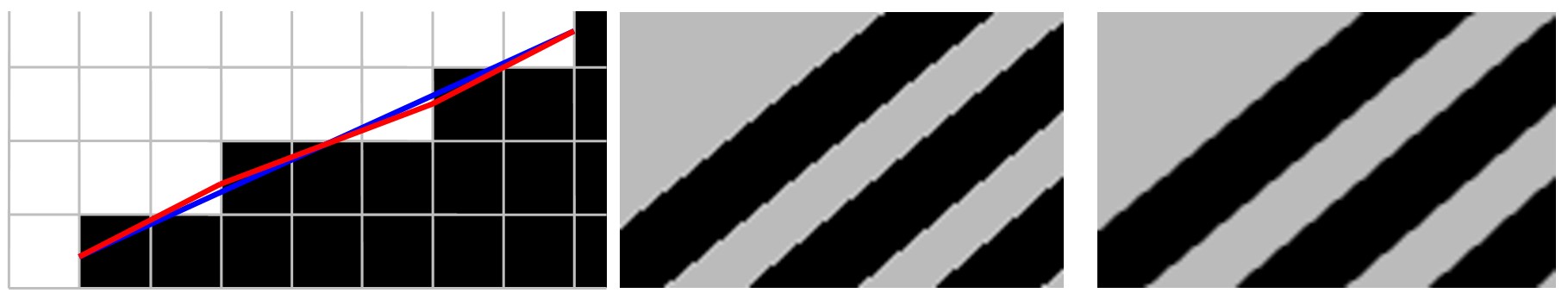}
    \caption{Left: the polyline is the reconstruction of the boundary by the morphological anti-aliasing algorithm; the straight line is the correction of the boundary reconstruction by the algorithm in this paper. Middle: the boundary without anti-aliasing processing. Right: the boundary line after the morphological anti-aliasing processing.}
    \label{fig1}
\end{figure*}

This paper finds that the main reason for the deviation of the reconstructed slanted line boundary from the original boundary in the morphological anti-aliasing algorithm is that these algorithms realize the reconstruction of the line boundary by detecting the jump of the line pixels in the horizontal or vertical direction. Due to the different step widths of pixel sawtooth generated by straight line boundaries with different slopes, there is a certain deviation between the reconstructed boundary based only on the local line shape and the real image. As shown in Figure 1, the shape
The dynamic anti-aliasing algorithm reconstructs the line shape locally every time a step is detected. The algorithm divides a complete straight line (left, dashed line in Figure \ref{fig1}) into segments. In actual use, this error will cause the undulating effect of the straight line boundary after anti-aliasing (right in Figure \ref{fig1}). If you want to obtain more complete global information of the straight line, the pixel search strategy using these algorithms will bring a high computational cost (see Section 2 for a detailed analysis).

\section{Related Works}

Aliasing is a common problem in both offline rendering \cite{wang2013gpu,wangimplementation,huo2015matrix,huo2016adaptive,huo2020spherical,cho2021weakly,fan2021real,huo2021survey,huo2020adaptive,huo2022extension} and real-time rendering \cite{kim2020single,li2021multi,an2021hypergraph,park2021meshchain,zhang2021powernet,li2020automatic}. Traditional anti-aliasing algorithms, supersampling anti-aliasing and multi-sampling anti-aliasing algorithms, have been the standard for anti-aliasing algorithms for more than ten years. The average value of the sampled colors. The idea of multi-sampling anti-aliasing is similar to that of super-sampling, and is usually considered a special case of super-sampling anti-aliasing. According to the definition of OpenGL \cite{r7}, multi-sampling anti-aliasing refers to An optimization of supersampling anti-aliasing, which generally means that some components of the final image are not oversampled, such as only supersampling the depth buffer and stencil buffer as the final color of this pixel. However, due to rendering efficiency and resources Many deficiencies in occupancy, coupled with the conflict with the current popular deferred rendering, have been unable to meet the requirements of the times. Although some recent related technologies, such as CSAA \cite{r8} and EQAA \cite{r9}, separate the color, depth and sample by The consumption of bandwidth and storage is reduced, but these methods still cannot avoid the shortcomings of oversampling and anti-aliasing.

The filter-based anti-aliasing algorithm of Yang et al. \cite{r10} is a very important progress in the field of anti-aliasing. Their directed adjustable boundary anti-aliasing filter is a GPU-based implementation method, which achieves super-sampling anti-aliasing. The grayscale variation level of aliasing. Their approach is to use the length of the contour line passing through a pixel as the weight of the sub-pixel. This approach can achieve the effect that the traditional approach of 2~3 times the sampling number can achieve. This method It is believed that it is not necessary to perform multi-sampling anti-aliasing on each pixel in the image, and the most serious anti-aliasing generally only occurs near the geometric boundaries, so it is only necessary to perform anti-aliasing processing at these boundaries. The algorithm still uses the traditional multi-sampling anti-aliasing algorithm. Although its final rendering effect is not superior to the original anti-aliasing algorithm, the idea of its filter points out another worthy improvement for the anti-aliasing algorithm. direction.

Reshetov's Morphological Anti-Aliasing (MLAA) \cite{r2} also proved that on the frame buffer without multi-sampling, very good anti-aliasing results can be obtained by boundary reconstruction and area calculation. Morphological anti-aliasing is a kind of post-processing. To deal with the anti-aliasing algorithm, it is necessary to reconstruct the information lost in the rendering process due to insufficient sampling. In short, the algorithm divides the local boundaries of the image into U-shaped, L-shaped and Z-shaped, and then infers the boundary according to different line types. Original morphology. In theory, morphological anti-aliasing can be divided into the following three steps: (1) boundary detection; (2) calculation of pixel mixing weights; (3) mixing with adjacent pixels. After Reshetov's morphological anti-aliasing algorithm, appeared There are many morphological anti-aliasing algorithms, but they basically follow this framework, including the algorithm in this paper.

Reshetov's morphological anti-aliasing algorithm is CPU-based and lacks practicability, which has prompted many researchers to propose GPU-based implementation methods, including Jimenez \cite{r10}, Biri \cite{r11}, AMD \cite{r12} and so on. In addition, the Fast Approximate Anti-Aliasing (FXAA) \cite{r3} algorithm proposed by Timothy of NVIDIA takes full advantage of hardware resources, for example, Fast Approximate Anti-Aliasing uses a non-homogeneous graph to estimate the end point of the line segment during the line segment search process. However, none of the above algorithms have proposed a solution to the anti-aliasing problem of sloping straight-line boundaries. Jimenez's improved sub-pixel morphological anti-aliasing (SMAA) \cite{r1} is also a morphological anti-aliasing algorithm based on GPU. Reference \cite{r1} For the first time, a processing method for oblique patterns is proposed, and the oblique boundary is anti-aliased to a certain extent. However, the algorithm of Jimenez et al. only considers the anti-aliasing of the boundary of the oblique line close to the angle of 45 degree, and ignores other slopes. 

When the image is not multisampled, the loss of some sub-pixel features is inevitable. In order to better reconstruct the original shape of the pattern, some researchers have also adopted the practice of recording necessary information in the rendering. Based on the morphological anti-aliasing algorithm, a buffer area that can save sub-pixel features is added (SRAA) \cite{r6}. Malan proposed the distance-to-boundary anti-aliasing method (DEAA) \cite{r4}, in the rendering process, it will calculate the distance from the pixel center to the triangle boundary, and store this information in the buffer area, and finally get the weight of the blend according to these distances. Similar to the distance-to-edge anti-aliasing algorithm, there is also Persson The geometric post-processing anti-aliasing (GPAA) \cite{r13} algorithm, etc. Through the additional information obtained in the rendering process, these algorithms can increase the accuracy of anti-aliasing, and to a certain extent, better deal with the aliasing of sloping straight-line boundaries However, recording additional data during the rendering process requires additional cache. In addition, this design method also makes the program lose the original morphological anti-aliasing post-processing characteristics, which will undoubtedly bring unnecessary use to the practical application of the algorithm. difficulty.

This paper is a post-processing morphological anti-aliasing algorithm based on aliasing graphics implemented by GPU. It follows the main ideas of boundary detection and color mixing of morphological anti-aliasing algorithm, and improves the linear boundary reconstruction in the process of weight calculation. The straight line is closer to the real straight line in the whole world. The algorithm in this paper reconstructs the boundary information of the straight line more accurately under the premise of adding very little extra calculation.

\section{Morphological Anti-aliasing Method for Boundary Slope Prediction}

\subsection{Algorithm Overview}

Morphological Anti-Aliasing (MLAA) \cite{r2} attempts to estimate the area covered by the original geometry in pixels. To accurately rasterize a triangle, it is necessary to obtain the area covered by each triangle in pixels. At the first rendering, the is a cached image without any anti-aliasing. To estimate the coverage area, the algorithm needs to re-vectorize the boundaries of the image in the cache.

The algorithm in this paper follows the basic framework of the morphological anti-aliasing algorithm, which is divided into three steps: (1) boundary detection; (2) area (mixing weight) calculation; (3) color mixing. The following briefly introduces boundary detection and color Mixed content, the specific algorithm details can be found in the literature [2]; after that, the core step of this paper in the morphological anti-aliasing algorithm - the improvement of the weight calculation will be explained in detail.

Boundary detection is the first step of the morphological anti-aliasing algorithm. If no boundary is detected, anti-aliasing processing will not be performed. There is a lot of information in the image that can be used for boundary detection, including color value, depth value, normal vector value, etc. Etc. This paper mainly determines whether the boundary exists by comparing the Luma values of two adjacent pixels. When the difference between the Luma values of two adjacent pixels is greater than a certain threshold, the boundary is considered to exist. For a pixel, it is only necessary to detect the above The information of border and left border, right border and bottom border can be obtained from the corresponding right and bottom pixels. For details about the algorithm of border detection, please refer to the literature \cite{r1}.

\subsection{Weight Calculation and Boundary Extension}

\begin{figure*}
    \centering
    \includegraphics[width=0.5\textwidth]{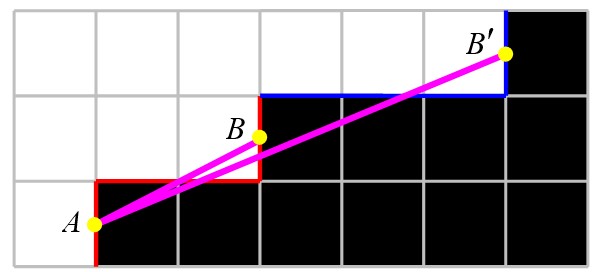}
    \caption{Line segment AB is the line segment reconstruction result of the original traditional morphological anti-aliasing algorithm. The goal of this paper is to correct point B to point B'.}
    \label{fig2}
\end{figure*}

The traditional morphological anti-aliasing algorithm divides the anti-aliasing calculation of the boundary into two directions—horizontal and vertical, which are carried out separately. Each boundary line segment is defined by two endpoints. To correct the incorrect line segment is essentially to find a more accurate line segment. The endpoints of the line segment. The traditional morphological anti-aliasing algorithm will assume that the endpoints of the line segment are A and B when encountering a Z-shaped (Figure \ref{fig2}, dashed line) boundary, however, A and B are probably not the true endpoints of the line .This detection method cannot accurately reflect the shape of the oblique straight line boundary. Using it for anti-aliasing calculation may cause the final rendering result to show a undulating effect in the oblique direction (Figure \ref{fig1}, right).

The Improved Subpixel Morphological Anti-aliasing Algorithm (SMAA) \cite{r1} proposes an anti-aliasing method to solve the oblique direction. The algorithm is inspired by the processing method of orthogonal patterns in the traditional morphological anti-aliasing algorithm. However, their algorithm can only deal with oblique straight lines around 45 degree. This paper proposes a more efficient and comprehensive solution. We believe that a section of boundary that appears in boundary detection is likely to be a A part of a longer straight line boundary. The algorithm in this paper corrects the incorrect estimation of the boundary in morphological anti-aliasing, making it closer to the real boundary.

The following is an example of how to correct the position of point $B$ to describe the morphological anti-aliasing method for boundary slope prediction. As shown in Figure \ref{fig2}, we need to correct the endpoint $B$ to the position $B'$. If the 3 pixels on the dotted line are traversed in turn , when the width of the step increases, the cost of the algorithm will be very high. The method in this paper is to directly predict the position of point B', and detect whether the boundary pattern of the position of point B' is the same as the boundary pattern of the original position of point B, if it is the same, It is considered that the right endpoint of this boundary is more likely to be $B'$ than $B$.

According to the Bresenham line drawing algorithm \cite{r14}, a line with a slope less than 45 degree is rasterized on the screen after the y of two horizontally adjacent pixels The coordinates remain the same or differ less than 1. To predict the position of B', the following theorem is introduced in this paper:

\begin{figure*}
    \centering
    \includegraphics[width=0.5\textwidth]{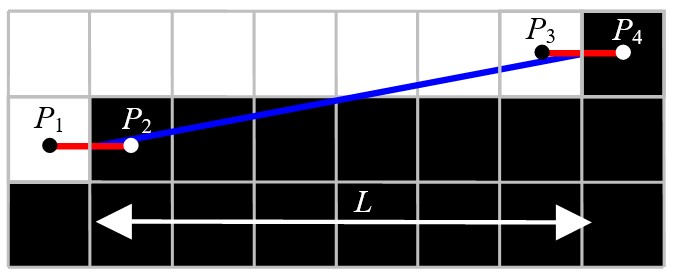}
    \caption{The width of the step is L, then the left and right endpoints of the line segment must be limited to the area of the $P_1 P_2$, $P_3 P_4$ line segments.}
    \label{fig3}
\end{figure*}

Theorem 1. For a straight line boundary, the width of each boundary pixel step does not differ by more than 1.

Proof: Prove by contradiction. First, let the width of the ladder be $L$, as shown in Figure \ref{fig3}, the left endpoint of the line segment is between $P_1$ and $P_2$, and the right endpoint of the line segment is between $P_3$ and $P_4$, then the slope of the ladder $k$ Satisfy the following inequality:

\begin{equation}
    \frac{1}{L+1}<k<\frac{1}{L-1}.
\end{equation}

Suppose the widths of the three steps on a straight line are $L_1$, $L_2$ and $L_3$ respectively, and they all satisfy the above inequalities.
\begin{equation}
\begin{aligned}
\frac{1}{L_1+1}<k<\frac{1}{L_1-1},\\
\frac{1}{L_2+1}<k<\frac{1}{L_2-1},\\
\frac{1}{L_3+1}<k<\frac{1}{L_3-1}.\\
\end{aligned}
\end{equation}

Let us assume $L_1<L_2$, that is $L_1+1 \leq L_2 \leq L_1-1$ , take the reciprocal of it to get: $\frac{1}{L_1-1}<\frac{1}{L_2}<\frac{1}{L_1+1}$, substituting the right-hand side of the inequality into the first inequality and substituting the left side into the third inequality, we get:

\begin{equation}
    \frac{1}{L_2}<k<\frac{1}{L_2},
\end{equation}
there is no suck $k$ exists.

Then according to Theorem 1, after the step width of a step is $L$ in the boundary detection, if the step is part of a straight line, the width of the next step can only be $L-1$ or $L+1$. In other words, if the step widths of the three sawtooths are $L$, $L-1$ and $L+1$ respectively, then the three sawtooths are definitely not on a straight line.

\begin{figure*}
    \centering
    \includegraphics[width=0.7\textwidth]{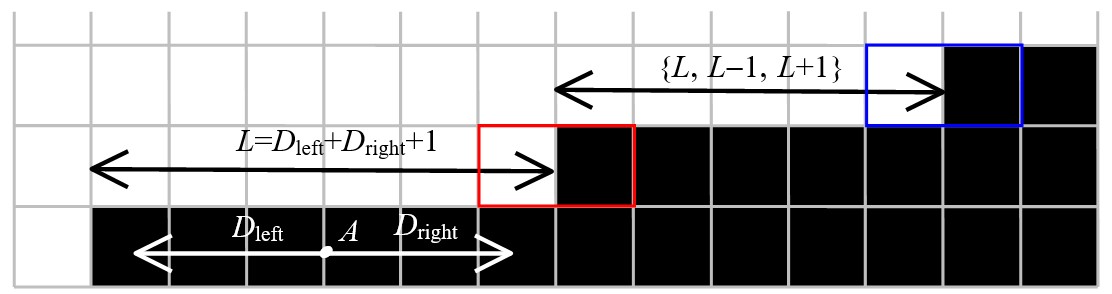}
    \caption{}
    \label{fig4}
\end{figure*}

\textbf{Implementation.} Assuming that there is an edge on the upper or lower boundary of a pixel (the edge mentioned here is the edge at the pixel level, not the boundary of the real object), you first need to find its left and right endpoints along this edge. The distances are assumed to be $D_{left}$ and $D_{right}$ respectively (as shown in Figure \ref{fig4}). Similarly, if there is an edge on the left or right edge of the pixel, it is also necessary to find the distance to the upper and lower endpoints along this edge, says $D_{up}$ and $D_{down}$ respectively. Point $A$ in the figure is the position of the current pixel; find the distances $D_{left}$ and $D_{right}$ to the left and right ends in the horizontal direction, the step width of this ladder is $L=D_{left}+D_{right}+1$; predict the next The step width is $L$, $L-1$ or $L+1$; compare the linetype at the new possible endpoint position (upper right box) with the linetype at the original endpoint (middle box).

\begin{figure*}
    \centering
    \includegraphics[width=0.2\textwidth]{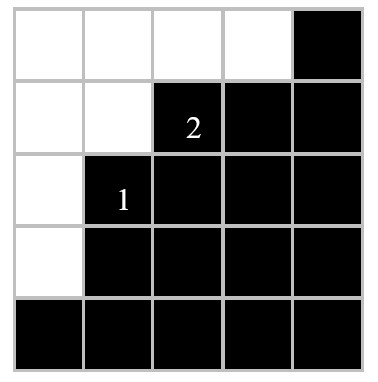}
    \caption{Pixel boundaries near position 1 tend to be vertical, and pixel boundaries near position 2 tend to be horizontal.}
    \label{fig5}
\end{figure*}

Next, it is necessary to compare the sizes of $D_{up}+D_{down}$ and $D_{left}+D_{right}$. If $D_{up}+D_{down}$ is larger, it can be judged that the distribution of the boundary around this pixel tends to be vertical (position 1 in Figure \ref{fig5}); if $D_{left}+D_{right}$ is larger, it can be considered that the border around this pixel tends to be horizontal (position 2 in Figure \ref{fig5}).

Assuming that the line type is mainly in the horizontal direction, it is necessary to first calculate its width $L=D_{left}+D_{right}+1$, and then extend the line to the left and right respectively, trying to find a more accurate endpoint. Take right search as an example, based on the previous According to the theory, the next step width may be $L$, $L-1$, $L+1$. Check whether the boundary information of these three positions is the same as the line type at the previous right endpoint (the box in Figure \ref{fig4}). Of course, due to the limitation of computing power, an upper limit N of the number of cycles needs to be set in advance. If the three positions $L$, $L-1$, $L+1$ are not linearly consistent with the boundary of the original endpoint position, then It is considered that the step is not part of the long straight line, and the search ends. If the new step width is detected as $L'$($L$ or $L-1$), it can be known that the next step width can only be $L$ or $L''$.

The pseudo code of the program is as follows:
\begin{algorithm}
\begin{algorithmic}
\State x=p
\For {i=0 to the upper limit of the number of loops N}
\If {pixel position x+L is the endpoint of the line segment}
    \State x=x+L
    \State keep looping
\EndIf
\If {the pixel position x+L-1 is the endpoint of the line segment}
    \State x=x+L-1,L'=L-1,
    \State jump out of the loop
\EndIf
\If {pixel position x+L is the endpoint of the line segment}
    \State x=x+1,L'=L+1,
    \State jump out of the loop
\EndIf
\State return x
\EndFor
\For {j=0 to N-i}
\If {pixel position x+L is the endpoint of the line segment}
\State x=x+L
\State keep looping
\EndIf
\If {the pixel position x+L' is the endpoint of the line segment}
\State x=x+L'
\State keep looping
\EndIf
\State return x
\EndFor
\end{algorithmic}
\end{algorithm}

The line type tends to be similar to the calculation in the vertical direction and the horizontal direction, and will not be repeated here.

The first step of the code initializes the current pixel position $x$ to the pixel $p$ that needs to calculate anti-aliasing. In the first loop, check whether the pixels of the three step widths ${L, L-1, L+1}$ are the boundary endpoints in turn, when it is found that the width of the end position is $L$, continue the cycle; when the width of the end position is found to be $L-1$, then the step width of the line is determined to be $L$ or $L-1$, so we assign $L-1$ to $L'$, and jump out Loop; when it is found that the width of the endpoint position is $L+1$, then the step width of the line is determined to be $L$ or $L+1$, so we assign $L+1$ to $L'$, and jump out of the loop; when it is found that none of the three adjacent pixels are When the boundary endpoint is reached, end endpoint detection and return the current endpoint position $x$.

After confirming that the straight line step has two widths of $L$ and $L'$, the program enters the second loop to detect whether the pixels at the two step widths of $L$ or $L'$ are the endpoints of the line segment. If it is the endpoint, continue the loop, if not, then end the endpoint detection and return the current endpoint position $x$.

\textbf{Algorithm Complexity Analysis.} It can be seen from the above pseudo code that in the first cycle, the detection of 3 endpoints is performed at most once, and in the second cycle, the detection of endpoints is performed at most 2 times. For endpoint detection, we use the morphological method in the literature \cite{r1}. For the method of boundary detection, one texture sampling can obtain the boundary endpoint information of two adjacent pixels. Therefore, the maximum number of sampling times required by our method is $N+2$, and the algorithm complexity is $O(N)$. If the proposed method in this paper is not used In the method of endpoint detection based on slope prediction, the edge-by-pixel endpoint detection method is adopted. Similarly, the texture calculation is performed once every two pixels along the step boundary. When the average step width of the line segment is $L$, a total of $L/2\cdot N$ sampling, the algorithm complexity is $O(L\cdot N)$. From the comparison of algorithm complexity, the algorithm proposed in this paper has better algorithm efficiency.

\section{Results}
In all the results, the result graph rendering in this paper adopts the upper limit of 4 cycles, and the sub-pixel reduction morphological anti-aliasing algorithm adopts the pixel anti-aliasing configuration of SMAA1x \cite{r1}. For each additional cycle, our method Search forward one time in both directions of the straight line. On the current experimental platform, the additional cost of adding one cycle is 0.07ms.

\begin{figure*}
    \centering
    \includegraphics[width=0.7\textwidth]{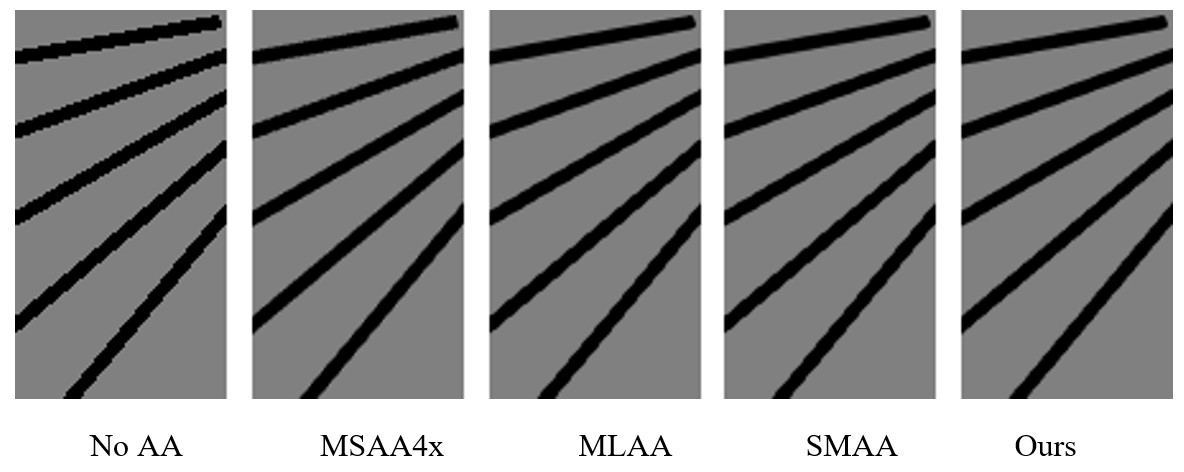}
    \caption{Comparison of Boundary Anti-Aliasing Results with Different Slopes.}
    \label{fig6}
\end{figure*}

\begin{figure*}
    \centering
    \includegraphics[width=0.7\textwidth]{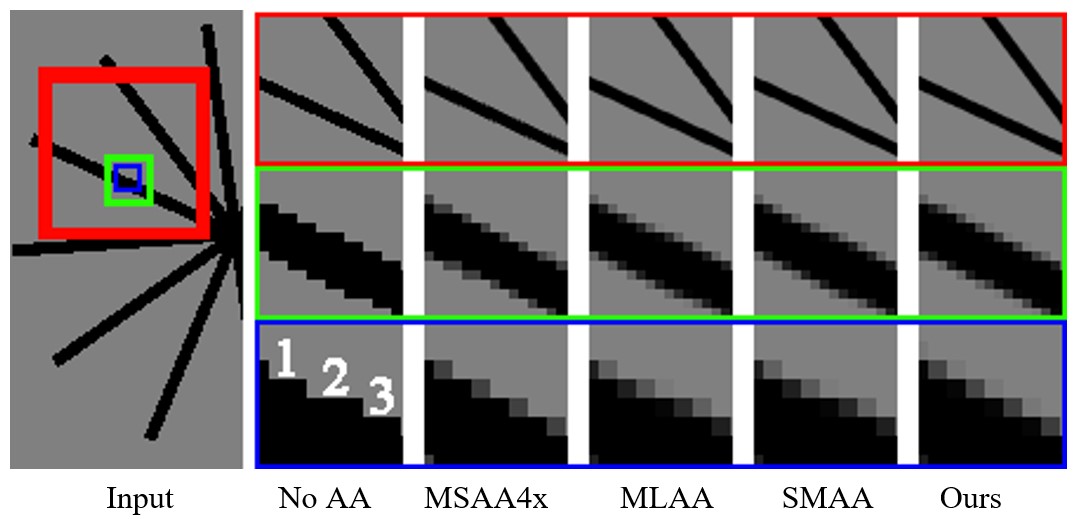}
    \caption{The big picture on the left is the original unprocessed picture, and the upper and lower right are the anti-aliasing in the area of the small box, the middle box and the large box corresponding to the original picture. As a result, the upper box area in the right image has not been enlarged, and the middle and lower box areas are the results of magnification by 4 times and 8 times, respectively.}
    \label{fig7}
\end{figure*}

\begin{figure*}
    \centering
    \includegraphics[width=0.9\textwidth]{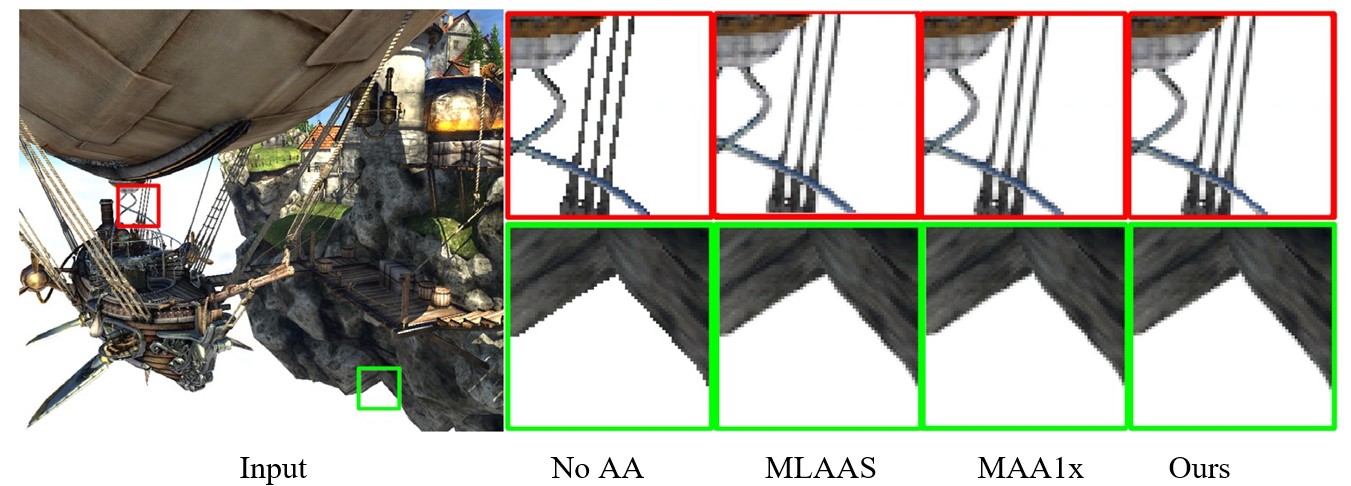}
    \caption{Anti-aliasing results of complex scenes (1), the original input comes from the literature \protect\cite{r1}.}
    \label{fig8}
\end{figure*}

\begin{figure*}
    \centering
    \includegraphics[width=0.9\textwidth]{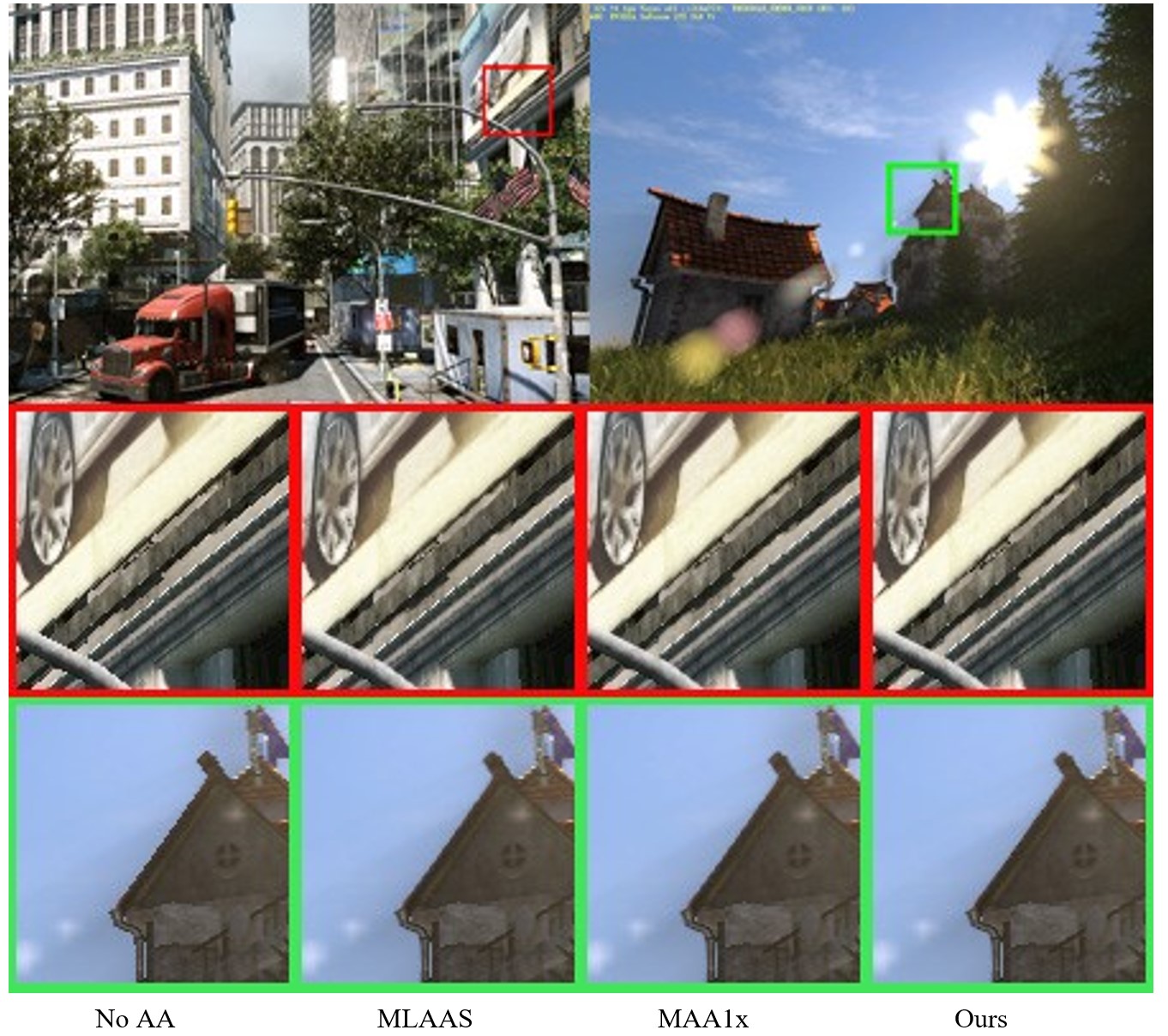}
    \caption{Anti-aliasing results of complex scenes (2), the original input comes from the literature \protect\cite{r1}.}
    \label{fig9}
\end{figure*}

We verified the algorithm proposed in this paper on multiple scenarios. Since the morphological anti-aliasing algorithm is an image post-processing method, we show the results in Figure \ref{fig6} and Figure \ref{fig7}.
Figure \ref{fig8} and Figure \ref{fig9} illustrate the result of directly importing the drawing image of a complex scene from the file as the original image, and performing anti-aliasing processing on the image. Since the operation of the algorithm in this paper is based on pixels, readers are advised to enlarge the picture for comparison.

Figure \ref{fig6} shows the anti-aliasing results of drawing a 3D scene. The scene is 5 cylinders with different inclination angles. We compare the multi-sampling anti-aliasing algorithm (MSAA4x), morphological anti-aliasing algorithm (MLAA), sub-pixel reduction morphological anti-aliasing algorithm (SMAA1x) and the results of the algorithm in this paper. From the boundary of the cylinder in Figure \ref{fig6} can be seen. The rendering results of the morphological anti-aliasing algorithm (MLAA) and the sub-pixel reduction morphological anti-aliasing algorithm (SMAA1x) have obvious fluctuations at the boundary of the line. The algorithm in this paper is based on the slope of the line boundary. Fixed, improved rendering to a large extent.

To further compare the above algorithms, we show a partial enlarged view of a 3D scene similar to the scene in Figure \ref{fig6}. As shown in Figure \ref{fig7} in the upper frame. The middle box and the lower box are the display results of 1x, 4x and 8x respectively. It can be seen from the leftmost 8x magnified image without anti-aliasing that the straight line boundary The width of the steps of the staircase varies between 2 pixels and 3 pixels, where the step width at position 2 is 3 pixels. Observe the position 1 and position in the diagram 3. It can be found that the pixel color of other algorithms is slightly lighter at position 1, and the pixel color at position 3 is slightly darker, which causes the rendering result to fluctuate at position 2. The reason. Through the correction of this article, the color of position 1 is darkened, and the color of position 3 is lighter, so that the straight line boundary appears smoother.

Figure \ref{fig8} and Figure \ref{fig9} show the anti-aliasing effect of the algorithm in this paper for the complex scene inclined straight line boundary. The display result in the box is the enlarged display result of the corresponding position in the figure. As can be seen from Figure \ref{fig8}, in the rock On the anti-aliasing of the lower boundary of the rope and the edge of the rope, the results obtained by our method show
Figure \ref{fig9} shows the comparison of the anti-aliasing results of the other two complex scenes. Comparing the anti-aliasing effects of the lower boundary of the billboard and the roof boundary on the left side of the house by different methods, the algorithm in this paper The processing result is also smoother and more natural.

\section{Conclusion}
Aiming at the shortcomings of morphological anti-aliasing, this paper proposes a method of using the slope of the straight line to correct the endpoints of the straight line accurately. The algorithm in this paper increases a small amount of computational cost and improves the deficiencies of morphological anti-aliasing on the boundary of the inclined straight line. A morphological anti-aliasing method based on the slope of the oblique straight line is proposed, which makes the anti-aliasing effect of the morphological anti-aliasing algorithm better, which will surely make the application of the morphological anti-aliasing algorithm further popularized.

\bibliographystyle{ieee}
\bibliography{srbib}

\end{document}